\newcommand{\be}{\begin{equation}}
\newcommand{\ee}{\end{equation}}
\newcommand{\bea}{\begin{eqnarray}}
\newcommand{\eea}{\end{eqnarray}}
\newcommand{\ba}{\begin{array}}
\newcommand{\ea}{\end{array}}
\documentclass{ws-mpla}

\begin{document}
\def\A{{\cal A}}\def\L{{\cal L}}\def\O{{\cal O}}
\def\la{\langle}\def\ra{\rangle}
\def\be{\begin{eqnarray}}\def\ee{\end{eqnarray}}
\def\lsim{\mathrel{\rlap{\lower3pt\hbox{\hskip1pt$\sim$}}
     \raise1pt\hbox{$<$}}} 
\def\gsim{\mathrel{\rlap{\lower3pt\hbox{\hskip1pt$\sim$}}
     \raise1pt\hbox{$>$}}} 
\def\bi{\bibitem}

\markboth{Hyun Kyu Lee, Mannque Rho and Sang-Jin Sin}
{Cold Dense Baryonic Matter and Compact Stars}

\catchline{}{}{}{}{}

\title{Cold Dense Baryonic Matter and Compact Stars \footnote{Based on the review prepared for the 2011 World Class University (WCU) International Conference, August 2011, Seoul, Korea.}
}

\author{\footnotesize Hyun Kyu Lee
}

\address{Department of Physics, Hanyang University\\
133-791 Seoul, Korea\\ hyunkyu@hanyang.ac.kr}

\author{Mannque Rho}

\address{Institut de Physique Th\'eorique,  CEA Saclay\\
91191 Gif-sur-Yvette C\'edex, France\\
 and Department of Physics, Hanyang University\\ 133-791 Seoul, Korea\\
mannque.rho@cea.fr}

\author{Sang-Jin Sin}

\address{Department of Physics, Hanyang University\\
133-791 Seoul, Korea\\ sjsin@hanyang.ac.kr}

\maketitle


\begin{abstract}

Probing dense hadronic matter is thus far an uncharted field of physics. Here
we give a brief summary of the highlights of what has been so far accomplished
and what will be done in the years ahead by the World Class University III Project at Hanyang University in the endeavor to unravel and elucidate the multifacet of the cold dense baryonic matter existing in the interior of the densest visible stable object in the Universe, i.e.,  neutron stars, strangeness stars and/or quark stars, from a modest and simplified starting point of an effective field theory modeled on the premise of QCD as well as from a gravity dual approach of hQCD. The core of the matter of our research is the possible origin of the $\sim 99\%$ of the proton mass that is to be accounted for and how the ``vacuum" can be tweaked so that the source of the mass generation can be uncovered by measurements made in terrestrial as well as space laboratories.  Some of the issues treated in the program concern what can be done -- both theoretically and experimentally -- in anticipation of what's to come for basic physics research in Korea.

\end{abstract}


\section{The Goal}
The proton which is understood to be made up of 3 ``chiral" quarks, 2 up quarks (denoted u) and 1 down quark (denoted d), is the most stable hadron with the
precisely measured mass $m_p=938.272013\pm 0.000023$ MeV. The quarks are light
fermions -- hence  chiral -- compared with the strong interaction scale $\sim
100$ MeV, with their mass $m_q=\frac 12 (m_u+m_d)\sim 4$ MeV. How the quark
masses (and also the lepton masses) arise, one of the most fundamental
questions of physics, will be clarified once the LHC finds the Higgs particle.
However the perhaps not less fundamental question is: Where does the bulk of
the proton mass come from?

This is a highly charged question. Take the nucleus which is the smallest
object that can be understood in terms of its constituents. The mass of a
nucleus of mass number $A$ can be expressed as
\be
M_A=Am_p+\delta \label{mass}
\ee
which is almost totally given by the sum of the proton mass with a small
binding energy correction $|\delta| \lsim 0.02 m_p$. This is the case with
everything that we see around us, atoms, molecules and buildings etc. Add the
mass of the constituents and that makes up the bulk. But when it comes to the
proton which is a constituent of a nucleus, this picture stops. It makes no
sense to write the proton mass as $m_p=3m_q+\Delta$. This is because what
corresponds to the ``correction" $\Delta$ amounts to 99\% of the proton mass
so the question is where do most  of the proton mass come from? This is one of
the most fundamental questions in physics, and an attempt to answer that
question leads all the way to asking what makes the densest object in the
Universe, neutron star, defy the gravitational collapse to a black hole. This
is the theme of the WCU3-Hanyang program.

In addressing this question, we assume our interpretation of what quantum chromodynamics (QCD) is telling us is correct: That most of the proton mass -- and other hadrons'mass made up of u and d (chiral) quarks -- are generated dynamically through spontaneous breaking of chiral symmetry, in other words, a sort of vacuum rearrangement. The objective of the WCU3-Hanyang program is how to ``see" the making of this mass by tweaking the ``vacuum structure."

The questions we pose to address this problem are the following:
\begin{enumerate}
\item What happens to the hadron masses under extreme conditions such as
    high temperature and, in particular, high density that are believed to
    change the vacuum structure ?
\item How is the structure of nuclei and cold baryonic matter modified
    under extreme conditions?
\item How does the modified nuclear structure influence dense matter in
    compact stars?
\item What makes the neutron stars stable against gravitational collapse?

\end{enumerate}

From cosmological considerations, it is not difficult to understand that a
strongly interacting matter above certain temperature must be in the form of
quarks and gluons, and not in hadrons. This means that going up, the
temperature must ``melt" the hadronic system. Indeed this understanding has
been confirmed in relativistic heavy-ion collisions, where the relevant
temperature has
been determined to be $\sim 175$ MeV. Future experiments at LHC/CERN will map
out the phase structure of such hot (but not so dense) matter revealing what takes place at high temperature.
However the situation with high density at low temperature is totally
different. It is nearly completely unknown what happens to cold hadronic system when it is squeezed to a very high density, other than that one is observing
compact stars with an interior density estimated to be as high as or even
higher than, say, $\sim 10$ times, that of nuclear matter. This is because cold dense hadronic matter is a strongly correlated system  inaccessible to
theoretical tools relying on standard perturbation theory, and furthermore the only reliable non-perturbative approach to QCD, i.e., lattice QCD, cannot handle, because of the famous sign problem, the density regime involved. In this paper, we describe the approach initiated by the WCU3-Hanyang team, what (little) has been so far accomplished and what remains to be done in the future. The earlier part of this development was summarized in \cite{PR}.

\section{Brown-Rho Scaling}\label{BR}
The underlying framework in this project is the Brown-Rho scaling predicted in
1991 by one of the authors with G.E. Brown~\cite{BR91}. A basic assumption that goes into this prediction was that if one ignores the small quark mass (the
process known as  ``the chiral limit"), the entire mass of the chiral-quark hadrons is generated by the spontaneous breakdown of chiral symmetry. Then in the limit that the number of colors $N_c$ is taken to be large~\footnote{Taking the large $N_c$ limit is the
only known nonperturbative analytical tool available so far for QCD which has been found to be reliable in certain processes for which lattice QCD calculation could be used to check, e.g., in the quenched approximation.} for which one can justify doing tree-order calculations, the hadron masses scale in the environment of dense medium with density denoted $n$ as
\be
m^*_N/m_N\approx m^*_M/m_M \equiv \Phi (n)
\label{BRS1}
\ee
where the subscripts $N$ and $M$ stand for the nucleon (proton, neutron) and
light-quark mesons ($\rho$, $\omega$), respectively and the asterisk for in-medium quantity at a given density. Here $\Phi(n)$ is a scaling function that depends on density $n$. The property of this scaling function is a subtle issue that is not precisely understood even now after two decades since its formulation in \cite{BR91}. The reasoning made there is based on the notion that the unbreaking of chiral symmetry is locked to the unbreaking of spontaneously broken scale symmetry in QCD. The underlying assumption that goes into (\ref{BRS1}) which was made more precise in \cite{LR} is that the scale symmetry breaking in QCD as manifested in the trace anomaly consists -- in the simplest description --  of two components, spontaneous breaking and explicit breaking, with the former associated with a ``soft gluon" and the latter with a ``hard gluon." What figures therefore in (\ref{BRS1}) is the soft component that can be described in terms of a scalar referred to as dilaton $\chi$ which is a Goldstone boson of the spontaneously broken scale symmetry. In terms of the vev of the dilaton in medium $\la\chi\ra^*\equiv \la\chi\ra_n$, one can associate the scaling function with the density dependence of the ``pion decay constant"\footnote{We put this quantity in the quotation mark since it is a parameter that appears in the Lagrangian which is related to the physical pion decay constant but can be identified with it only in the mean-field or tree approximation. We will make this distinction more precise later.}
\be
F_\pi^*=F_\pi\frac{\la\chi\ra_n}{\la\chi\ra_0}
\ee
as
\be
\Phi(n)=F_\pi^*/F_\pi.
\ee
It seems natural that this parameter appears on the RHS of Eq.(\ref{BRS1}) because at low density, it is the relevant parameter connected to the order parameter of chiral symmetry, the quark condensate $\la\bar{q}q\ra$. That it appears linearly is a consequence of the assumption taken. It could in principle be a more involved function of it. It is important to note that the quantities in (\ref{BRS1}) are the {\em intrinsic or bare parameters} of the Lagrangian defined at a given density $n$.  When $1/N_c$ corrections enter importantly via quantum loop graphs, the scaling of physical quantities such as masses will be functions of those quantities but not necessarily in the same simple form. This point is often misinterpreted in the literature causing confusion of what the evidence is for chiral symmetry restoration, partial or complete.

There are unmistakeable evidences, albeit indirect, that the scaling
(\ref{BRS1})  linear in $F_\pi^*$ does hold
at least qualitatively or even semi-quantitatively in nuclear medium up to nuclear matter density $n_0\simeq 0.16$ fm$^{-3}$. There is, however, up to date no ``smoking-gun" signal for its direct connection to chiral symmetry, namely, the quark condensate. It remains still a controversial issue. On the other hand, it is safe to say that contrary to what is sometimes claimed, there is no clear evidence against it either. We will return to this matter in a different context.

Naively one would think that since the (physical) pion decay constant is supposed to go to zero at high density where chiral symmetry is supposed to be restored, the mass would disappear as the pion decay constant goes to
zero. However there is no theoretical argument available at present to show
that the mass has to go to zero in the simple form of (\ref{BRS1}). Even if it
goes to zero, it could be a complicated function of $F_\pi^*$ that vanishes at the chiral restoration point. To avoid confusion, the physical pion decay constant that vanishes at chiral restoration will be denoted $f_\pi^*$. It turns out that if one implements nonlinear sigma model --
the established effective field theory at low energy -- with hidden local
symmetry (HLS for short) associated with the vector mesons that are known to be relevant degrees of freedom in low-energy strong interaction dynamics\footnote{An elegant and powerful way to incorporate the energy scale corresponding to the vector meson mass $\sim 800$ MeV in the non-linear sigma model is the hidden local symmetry approach of M. Harada and K. Yamawaki in Phys. Rept. {\bf 381},  1 (2003) who showed how to consistently do chiral perturbation calculation in the presence of vector mesons and approach chiral phase transitions at high temperature and/or high density. This hidden local symmetry appears naturally in the form of infinite tower in holographic dual QCD derived from string theory, e.g., T. Sakai and S. Sugimoto, {Prog. Theor. Phys.} {\bf 113}, 843 (2005).} one finds that the RHS of (\ref{BRS1}) has to be replaced as one approaches near the critical point -- $T_c$ or $n_c$ -- by
\be
m^*_N/m_N\approx m^*_M/m_M\approx \la\bar{q}q\ra^*/\la\bar{q}q\ra.
\label{BRS2}
\ee
This is a relation which should be valid {\em if at all} only very near the critical point with  modulo an overall constant of order 1.
There is at present no experimental confirmation or falsification of this
relation. There are several experimental papers that claim evidences either for or against it but we consider those claims to be unfounded. A compelling argument was given in \cite{BHHRS,mr-Gerryfest} that the experiments so far performed and analyzed by several model calculations {\em did not measure} the observables that are sensitive to chiral symmetry properties. The verdict is still not out, awaiting more precisely focused experiments and
interpretations thereof.

It should be stressed that the hidden local symmetric approach to the scaling (\ref{BRS2}) relies on several rather strong assumptions: (1) The density driven transition is a smooth cross-over or second-order or at worst weakly first-order; (2) the matching of the one-loop HLS correlators with the operator product expansion of QCD correlators at the ``chiral scale" $\sim 1$ GeV is reliable; and (3) the fermions, quasiquarks or baryons, in the effective Lagrangian are {\em the} relevant degrees of freedom. One or more of these assumptions can go wrong in several ways. There are some obvious caveats to these assumptions. Firstly, there can be a chiral-invariant mass term contributing to the nucleon or quasiquark mass, which would then invalidate both the LHS and RHS of (\ref{BRS2}). This possibility has been considered in terms of a parity-doublet hidden local symmetric model that allows a chiral-invariant mass term that remains non-zero in the chirally restored phase. This matter will be discussed below. Secondly in doing the matching of the correlators at a matching scale deemed optimal for both the QCD correlators and the HLS correlators,  higher-dimension chiral order parameters that can enter are expressed in terms of the quark condensate $\la\bar{q}q\ra$. But this cannot be rigorously justified when baryonic systems are simulated on crystals as discussed below. We will indeed find that at some density which will be labeled as $n_{1/2}$, $\la\bar{q}q\ra$ vanishes but the pion decay constant $f_\pi^*$ does not, signaling that the system is still in chiral symmetry-broken phase: The non-vanishing pion decay constant must be getting contributions from higher-dimension order parameters characterizing broken chiral symmetry. Thirdly, the large $N_c$ assumption which is the basis for the definition of the ``mass" could break down, with loop corrections becoming crucial as mentioned above.

Given that cold baryonic matter at high density defies QCD-inspired approaches, it is tempting to exploit the presumed power of gravity dual models anchored on string theory to address the density regime involved. Indeed such a strategy
has been found to be successful in certain processes in condensed matter and
also in relativistic heavy-ion collisions that are not accessible in
perturbation theory. The well-known case is the low shear-viscosity at high
temperature seen in RHIC experiments which cannot be explained in perturbative QCD but can be explained in terms of the gravity-dual of ${\cal N}=4$ super-symmetric theory although the latter is not QCD in the UV regime. This class of success  led to the attempt to apply a holographic QCD approach to in-medium meson properties in the difficult density regime~\cite{jrss}. Qualitatively the scaling (\ref{BRS1}) at low density seems to be reproduced, but the high density scaling (\ref{BRS2}) was not seen at all in the model. There can be a variety of reasons for this deviation from the ``gauge-theory" prediction. One mechanism  un-mistakenly identifiable in QCD-inspired models such as Nambu-Jona-Lasinio is the important role played by low-mass scalar mesons and $1/N_c$ corrections associated with them~\footnote{In such models, it is seen that if $N_c$ exceeds 3 found in nature, nuclei and nuclear matter cannot be bound. See L. Bonanno and F. Giacosa, arXiv:1102.3367[hep-ph]. What this implices in the bulk sector is not clear and deserves to be clarified.}, both of which are missing in  holographic QCD. Implementing such scalars in hQCD is an open problem.
\section{Half-Skyrmion Matter and New BR}
The first indication that (\ref{BRS2}) may break down at some density above $n_0$ in dense medium was noticed when soliton configurations in nonlinear sigma model describing nucleons as skyrmions are put on an FCC crystal lattice to simulate dense matter. At present, this is the only way we know how to simulate nonperturbatively dense matter using an effective Lagrangian. Lattice gauge theory is unable to handle the density involved and all model approaches start perturbatively from the matter-free vacuum with phenomenological Lagrangians. These may be made to work up to the nuclear matter density with the help of experimental data but cannot be trusted when extrapolated to higher density.

The crystal simulation that we employ can be justified in the large $N_c$ limit for baryonic matter at high density. What is intriguing in this approach is the observation that at some density $n_{1/2}> n_0$ (where $n_0\approx 0.16$ fm$^{-3}$ is the normal nuclear matter density), a skyrmion in the FCC configuration fractionizes into two half-skrymions, each carrying a half baryon charge in the BCC configuration. This feature shown in Fig.~1 is not observed in any other treatments\footnote{There is a resemblance of this phase to what is called ``vector symmetry" discussed by H. Georgi in {Phys.\ Rev.\ Lett.}\ {\bf
63}, 1917 (1989).}. This structure was also observed~\cite{rsz} with the instantons of the holographic QCD models that possess correct chiral symmetry structure put on an FCC lattice. Here what corresponds to the half-skyrmion is the dyon in the instanton structure of \cite{rsz}. Independently of detailed mechanisms involved, the import of this half-skyrmion-dyon identification is that the fractionized topological soliton could be a generic feature of the topological nature of the nucleon, and not exotic as has been largely thought. And it seems highly likely that it is not just an artifact of the crystal structure either. What matters is at what density the transition takes place.
\begin{figure}[h]
\centerline{\psfig{file=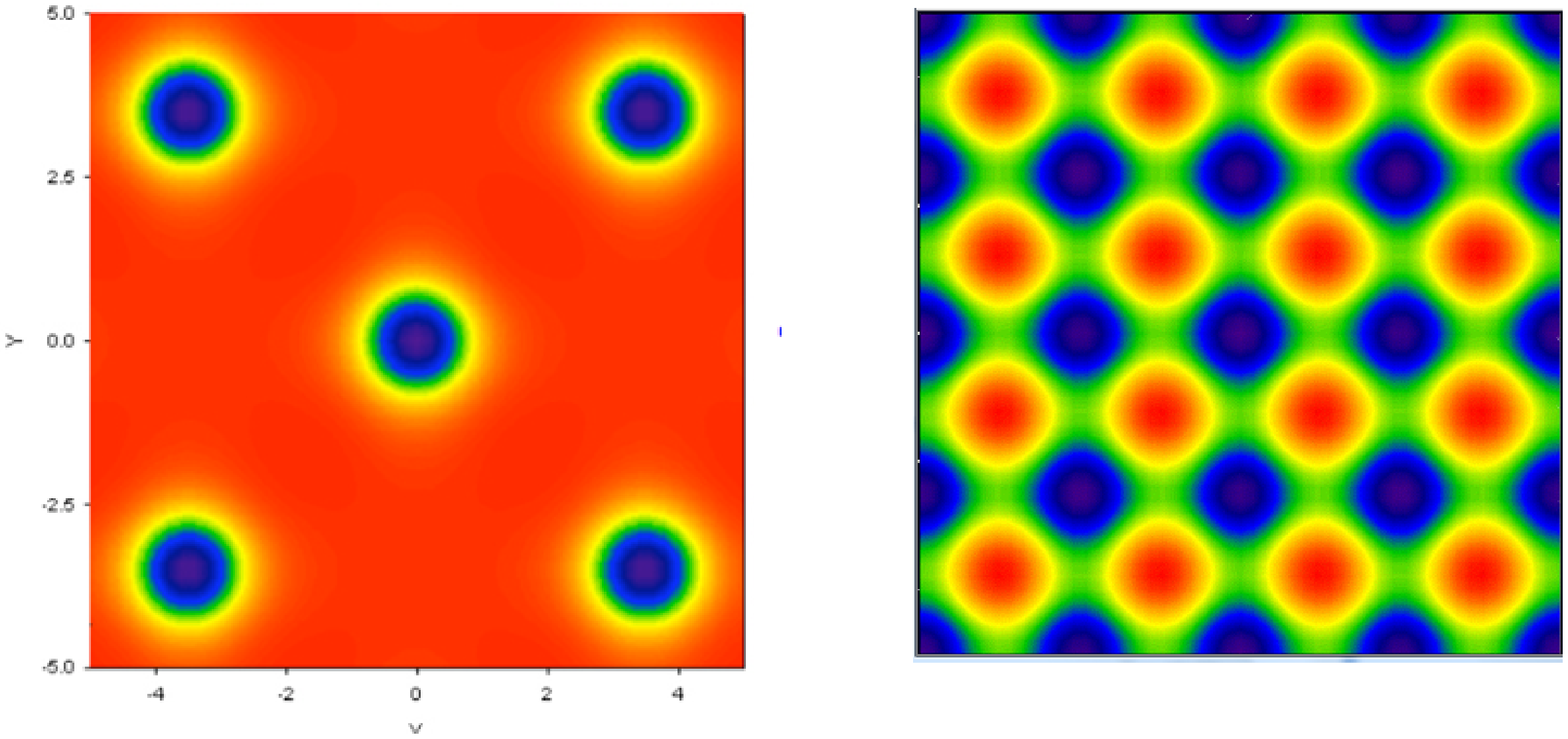,width=10cm,angle=-90}} \vskip -3.5cm
 \caption{(Color online) Local baryon number densities for low-density skyrmion phase (left
panel) in FCC and high-density half-skyrmion phase in BCC (right panel). The red
color represents the empty space. This is taken from
  Lee {\it et al.}, Nucl. Phys. {\bf A723}, 427 (2003).}
\label{density}
\end{figure}

What is distinctive of the half-skyrmion structure is that it arises as a sort
of phase transition at a density $n_{1/2}$ lying higher than
$n_0$, with the chiral condensate vanishing in the unit cell,
$\la\bar{q}q\ra=0$, but a non-vanishing pion decay constant $f_\pi^*$. Although the quark condensate -- which is the standard order parameter for chiral symmetry -- is zero, the non-vanishing pion decay constant says that chiral symmetry is not actually restored. This means that it cannot be a standard order parameter. This may be symptomatic of the crystal lattice structure, which may not support chiral symmetry restoration. However it is a phase transition in that there is change in the relevant degrees of freedom from skyrmions to half-skrymions, involving also a topology change. This resembles rather ``deconfined" quantum critical phenomenon in condensed matter physics which goes outside of the Landua-Ginzburg-Wilson paradigm of phase
transition.\footnote{See T. Senthil {\em et al.}\, Nature {\bf 302}, 1490
(2004).}

There are several striking consequences in nuclear dynamics of this
skyrmion-half-skyrmion phase transition, in particular in cold dense baryonic
matter. When a scalar degree of freedom is incorporated via a dilaton
associated with the spontaneously broken scale symmetry in QCD (in addition to
the explicit breaking due to quantum anomaly, i.e., the QCD trace anomaly) into
nonlinear sigma model as needed for low-energy nuclear dynamics~\cite{LR}, the
scaling (\ref{BRS1}) is found to hold fairly well up to
$n\approx n_{1/2}$, but above $n_{1/2}$, the scaling (\ref{BRS2}) gets
significantly modified. It takes the form~\cite{LPR}
\be
 m^*_M/m_M &\approx& \kappa (g^*/g), \label{vector}\\
 m^*_N/m_N &\approx& \kappa. \label{BRS3}
\ee
Here $g$ is the hidden gauge coupling constant and $\kappa$ is a constant
weakly dependent on density, given
in the skyrmion model as $F_\pi^*/F_\pi$.\footnote{The physical pion decay
constant in medium or more precisely the temporal pion decay constant given by
$f^{t*}_\pi=F^*_\pi+\delta F_\pi$ where $\delta F_\pi$ stands for loop
corrections, goes to zero -- in the chiral limit -- at the chiral restoration, but $F^*_\pi$ does not.} The hidden gauge coupling constant goes like $g\sim \la\bar{q}q\ra$ near the chiral restoration point, so in the vicinity of chiral restoration it is the hidden gauge coupling that takes over the role of chiral order parameter. Away from the critical point, how (\ref{vector}) scales is not known.  As a whole, this modification in the scaling makes a drastic change in the structure of nuclear forces. Most notable is the change in the nuclear tensor forces.
\section{Tensor Forces}\label{tensorforce}
The new scaling predicted by the topology change (\ref{vector}) and
(\ref{BRS3}) can have a dramatic effect in tensor forces in dense matter for $n>n_{1/2}$. It is easy to see what happens assuming that
tensor forces are mediated by the exchange of the pion and the $\rho$ meson.
Suppose that one can treat the nucleon as non-relativistic. Then the tensor
forces take the form
 \begin{eqnarray}
V_M^T(r)&&= S_M\frac{f_{NM}^2}{4\pi}m_M \tau_1 \cdot \tau_2 S_{12}\nonumber\\
&& \left(
 \left[ \frac{1}{(m_M r)^3} + \frac{1}{(m_M r)^2}
+ \frac{1}{3 m_Mr} \right] e^{-m_M r}\right),
\label{tenforce}
\end{eqnarray}
where $M=\pi, \rho$, $S_{\rho(\pi)}=+1(-1)$.
The total tensor force is then the sum $V^T=V_\pi^T +V_\rho^T$ and since they
come with an opposite sign, they tend to cancel.

\begin{figure}[ht!]
\begin{center}
\includegraphics[height=6.cm]{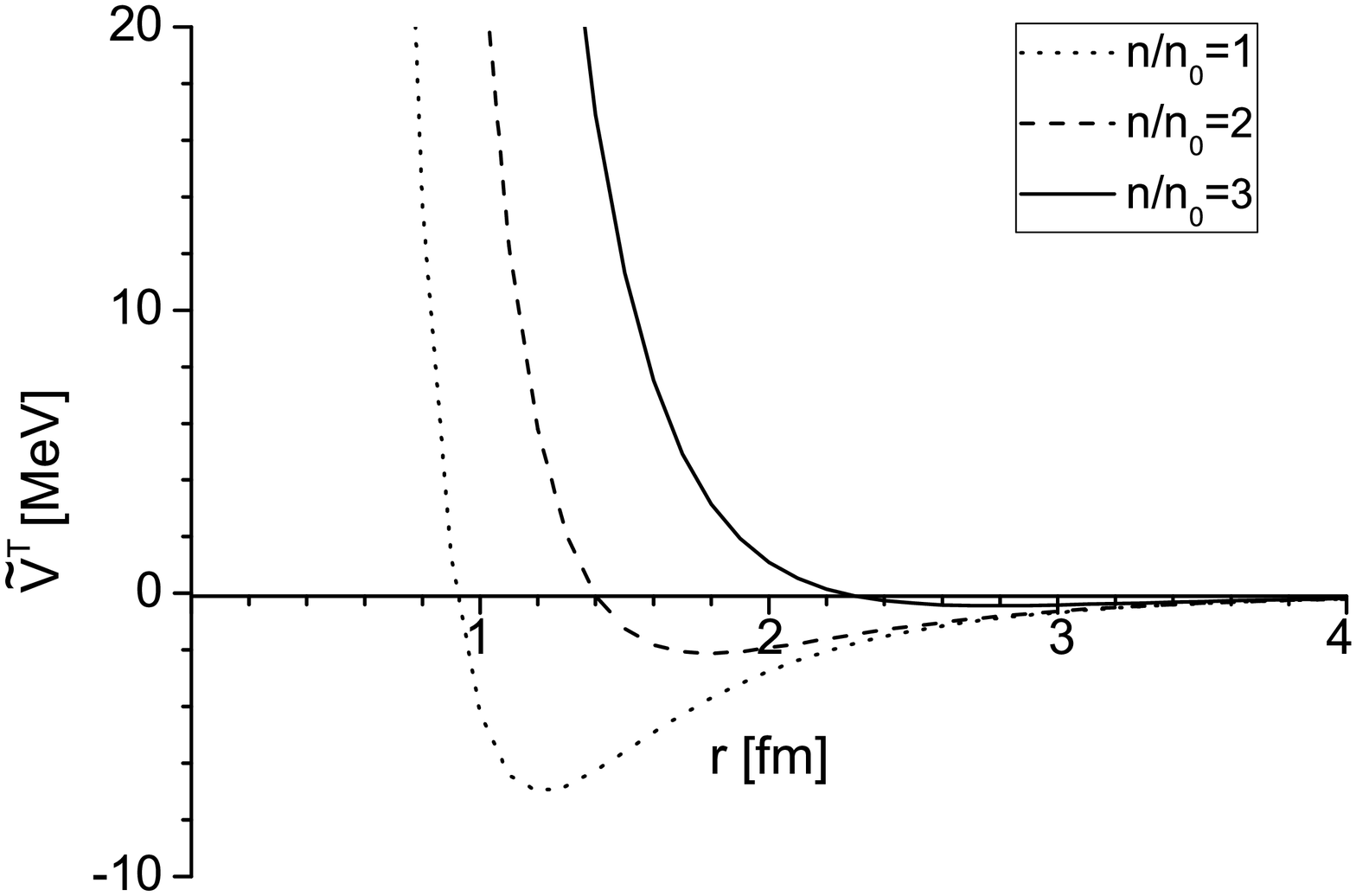}
\includegraphics[height=6.cm]{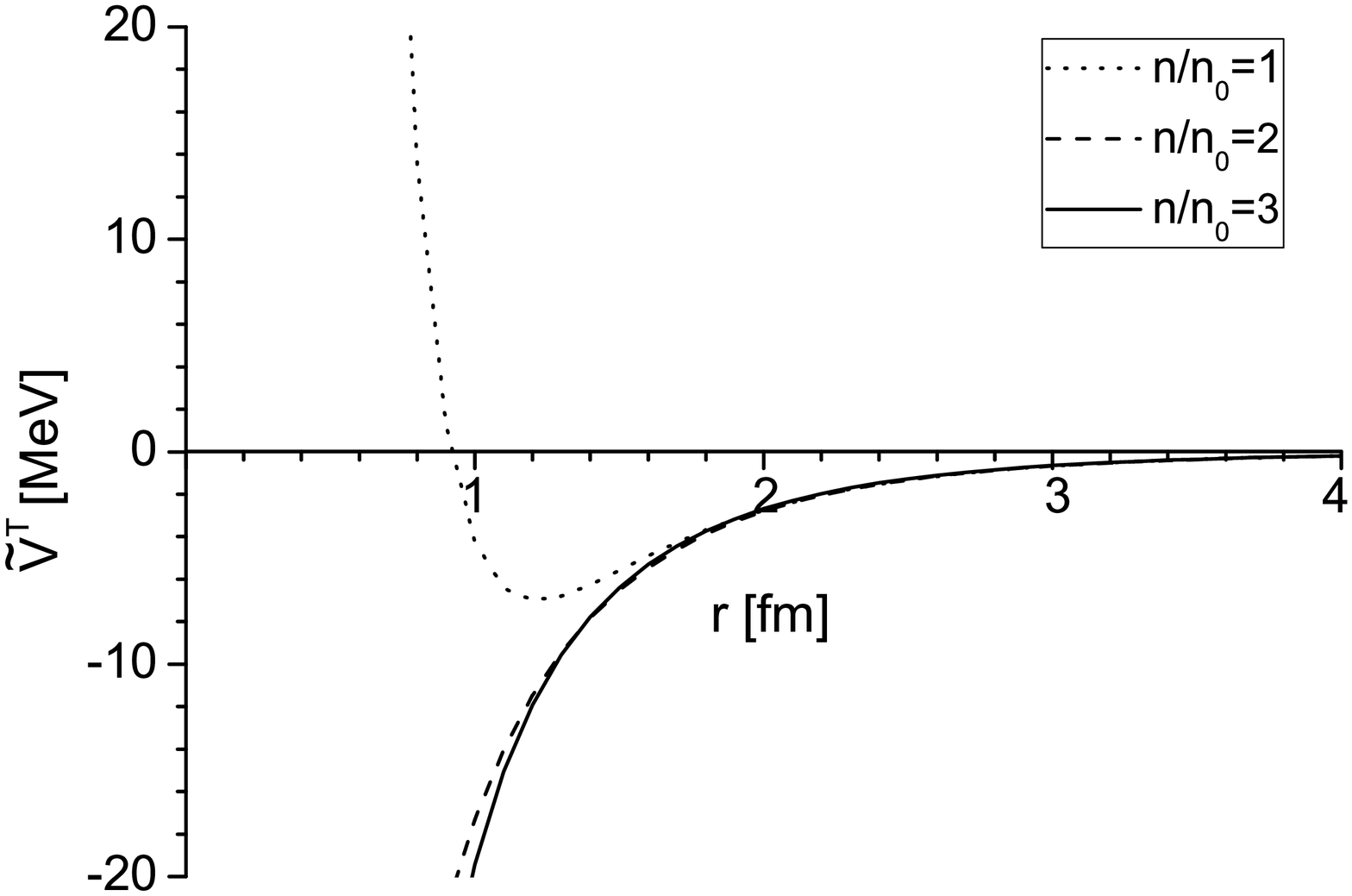}
\vskip -0.5cm
\caption{Sum of $\pi$ and $\rho$ tensor forces in units of MeV as function of
densities $n/n_0$ =1, 2 and 3 with the ``old scaling" (\ref{old}) (upper panel)
and ``new scaling" (\ref{new}) (lower panel).}\label{tensor}
\end{center}
\end{figure}

To exhibit the consequence of the scaling modified by the half-skyrmion phase,
it is convenient to express the old and new scalings as follows. \be
(m^*_N/m_N)_{old}\approx (m^*_M/m_M)_{old}\approx (f_\pi^*/f_\pi)=\Phi(n),\
(g^*/g)\approx 1\ \ {\rm for}\ 0\lsim n\lsim n_c\label{old}
\ee
and
\be
(m^*_N/m_N)_{new}&\approx& (m^*_M/m_M)_{new}\approx (f_\pi^*/f_\pi)=\Phi(n),\
(g^*/g)\approx 1\ \ {\rm for}\ 0\lsim n\lsim n_{/2}\ ,\nonumber\\
(m^*_M/m_M)_{new}&\approx& \kappa (g^*/g)\approx\Phi^\prime(n), \ \
(m^*_N/m_N)_{new}\approx \kappa \ \ {\rm for}\ n_{1/2}\lsim n\lsim
n_c\ .\label{new}
\ee
Up to $n\approx n_0$, the two scalings are the same and constrained by
experiments. Unless $n_{1/2}$ is much greater than $n_0$, it should be safe to
take the meson and baryon scalings to be the same up to the density $n_{1/2}$. Beyond $n_{1/2}$, however, the scalings $\Phi$ and $\Phi^\prime$ could differ. In fact, $\Phi^\prime$ is likely to fall faster than $\Phi$. In Fig.~\ref{tensor} is shown how the tensor forces would look like for a reasonable set of parameters~\cite{LPR}. While details depend on the parameters that we are unable to pin down precisely, the qualitative feature will remain unchanged. The new scaling predicts a tensor force structure drastically different from the old scaling above $n_{1/2}$. Specifically at $n\sim 3n_0$, the tensor force that results from the cancelation between the two is nearly zero with the old scaling, whereas with the new scaling, the $\rho$ tensor gets strongly suppressed, leaving only the pion tensor operative. This effectively increases the tensor force for $n\gsim n_{1/2}$.

\section{Effect of New BR on Symmetry Energy}
The new scaling will affect dramatically the symmetry energy factor $S$ in asymmetric nuclei, and of course neutron-star matter, defined by
\be
E(n,x)=E(n,0)+S(n)x^2+\cdots\label{SE}
\ee
where $x=(N-P)/A$ with $A=N+P$ with $N$ and $P$ standing for the neutron number and the proton number, respectively.  The factor $S$ has been measured up to $n\sim n_0$ in neutron-rich nuclei but there is little information on it for $n\gsim n_0$ which figures importantly for the EoS of compact stars. There is no reliable theoretical tool to determine $S$ at high density accounting for the widely diverging theoretical predictions in the literature.

It has been argued that at high densities, $S$ is dominated by the tensor forces. A rough but good enough approximation is
\be
S\propto |V^T|^2/\bar{E}\label{tensorcon}
\ee
where $\bar{E}$ is the average excitation energy associated with the tensor
force, $\bar{E}\sim 200$ MeV.  There is kinetic energy contribution to $S$ which is not in (\ref{tensorcon}) but it is known to be small, so we shall ignore it here. With the tensor force given by the old scaling, Fig.~\ref{tensor}, one would find that the tensor force contribution to $S$  nearly vanish at $n\gsim 2n_0$ . One might think that this feature agrees with what is called ``supersoft" symmetry energy that vanishes at $n\sim 3n_0$\footnote{An evidence for such a supersoft $S$ is discussed by Z.~Xiao et al.,   Phys.\ Rev.\ Lett.\  {\bf 102}, 062502 (2009).}.  However, the situation is totally different with the new scaling.
There the $\rho$ tensor gets strongly quenched for $n\gsim n_{1/2}$, leaving the pion tensor untouched, which means that $S$ will increase as the net tensor force gets stronger as density increases beyond $n_{1/2}$, making the symmetry energy {\em stiffer again}. We will see this feature supported by the dilaton-limit fixed point discovered recently as discussed in Section \ref{dialton}.  This phenomenon is expected to be highly relevant to confronting the recently measured 1.97 solar-mass neutron star. The Radioactive Ion Beam (RIB) machines -- such as, e.g., the KoRIA in project in Korea -- are expected to provide a valuable information on the symmetry energy $S$ just above $n_0$ where experimental data are lacking.
\section{The ``Ice-9" Phenomenon}
\begin{figure}[ht]  
\centerline{
\epsfig{file=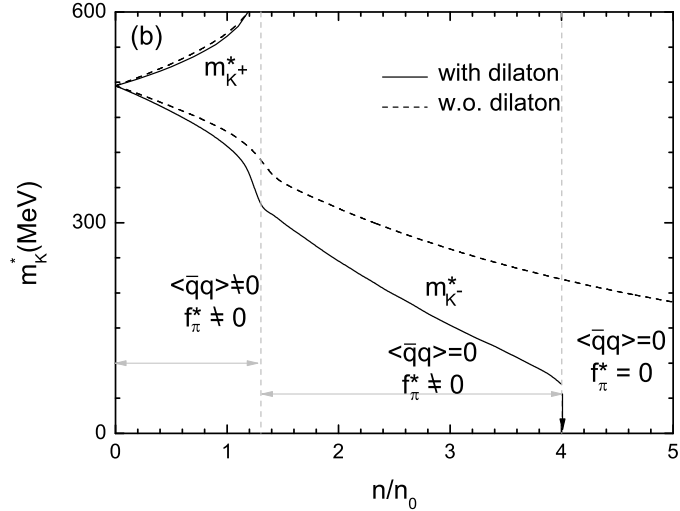, width=9cm, angle=0}}
\caption{ $m^*_{K^\pm}$ vs. $n/n_0$ (where $n_0\simeq 0.16$ fm$^{-3}$
is the normal nuclear matter density) in dense skyrmion matter
which consists of three phases: (a) $\la\bar{q}q\ra\neq 0$ and
$f_\pi^*\neq 0$, (b) $\la\bar{q}q\ra=0$ and $f_\pi^*\neq 0$ and
(c) $\la\bar{q}q\ra=0$ and $f_\pi^*=0$. The role of the dilaton associated with the spontaneous breaking of scale invariance described in Section \ref{BR} is indicated.
}
\label{ice9}
\end{figure}

Another potentially important effect of the half-skyrmion phase is on deeply bound dense kaonic nuclei. It manifests itself in the attraction felt by an anti-kaon embedded in dense matter. The phase change can induce an enhanced attraction\cite{PKR}, as shown in Fig.~\ref{ice9}, hitherto unseen in standard nuclear theory, at $n\sim n_{1/2}$ such that a nucleus with one or more bound anti-kaons could become ultra-compact with the average density exceeding that of ordinary nuclear matter. Such a compact nucleus, e.g., $K^-K^-pp$, could be produced in future experiments at GSI and/or J-Parc~\footnote{For a recent discussion, see T. Yamazaki {\it et al.}, arXiv:1106.3321[nucl-th].}. This phenomenon has also an implication on kaon condensation in compact-star matter discussed below.

The precipitous drop in the kaon mass at $n=n_{1/2}$ could trigger the
formation of deeply bound anti-kaon-nuclear state. In conventional treatments, i.e., under normal conditions, such a process cannot occur so it may be dubbed as  ``ice-9" phenomenon.\footnote{Gerry Brown suggested to us the analogy between the superdense kaonic nuclei and  Kurt Vonnegut's ``ice-9" in {\it Cat's Cradle}.}
\section{Three-Layer Compact Stars}\label{star}
The effect of the modification in scaling due to the topology change described
in Section \ref{tensorforce} is expected to influence not only the structure of finite nuclei and nuclear matter but also that of compact stars. How this modified BR scaling will affect the EoS of
neutron stars is being studied in a nuclear effective field theory approach anchored on $V_{lowk}$ which is derived via Wilsonian renormalization group equations from precision nucleon-nucleon interactions~\cite{Dongetal}. This work will help provide badly needed constraints on the poorly known symmetry energy at high density.

The presence of a half-skyrmion phase at a density not too high above $n_0$ affects not only the tensor forces but also the anti-kaon spectrum. The symmetry energy factor $S$ will also be affected by the possible role of the anti-kaons condensed into the system. It has been suggested that the recently discovered high-mass neutron star PSRJ1614-2230 (with mass of ($1.97\pm 0.04$) $M_\odot$) could rule out the possibility of hyperon and/or kaon condensation in compact star matter\footnote{See P.B. Demorest {\it et al.}, Nature {\bf 467}, 1081 (2010).}. The simple reason for this assertion is that such strangeness degrees of freedom would render the EoS too soft to resist the collapse to a black hole. In the work done recently\cite{KLR}, a scenario of matter composed of three phases, namely, nuclear matter at the outer layer, kaon-condensed matter in the intermediate layer and quark matter in the interior is developed into a three-layer model that could give an M vs. R relation compatible with the observation.

To inject strangeness in dense baryonic matter requires treating both the hyperon degrees of freedom and the K-meson degrees of freedom simultaneously. This is a highly nonlinear and intricate matter and it is not known how to do this in a consistent way. One possibility is to start with a three-flavor hidden local symmetric Lagrangian, generate octet baryons as solitons and then treat the baryons and octet mesons on the same footing in dense medium. We are working on this problem but we are at the initial stage and have no results to show. The approach followed in \cite{KLR} is to focus on the symmetry energy and consider the effect of strangeness in the symmetry energy in terms of the kaon degrees of freedom with the hyperon fields (put in by hand) in the $SU(3)$ chiral Lagrangian integrated out\footnote{This procedure will be invalidated if there is an infrared enhancement in kaon-hyperon interactions.}. The effect of the hyperons so integrated out would then be lodged in the density-dependent parameters of the kaon-nuclear Lagrangian that we will treat in the simple tree order with the multi-kaon interactions suppressed. There are indications from lattice simulations\footnote{See W. Detmold {\em et al}, Phys.Rev. D{\bf 78}, 054514 (2008).} that such multi-kaon interactions are not important although the effect of baryons in the multikaon interactions is neglected in the lattice calculations.

The key ingredient in our treatment for kaon condensation is the decrease of the effective mass denoted as $m_K^*$\footnote{In this paper,  $m^*_K$ is the kaon energy in medium since we are dealing with the s-wave kaon.} of the negatively charged kaon $K^-$ as density increases.  The $m_K^*$ is basically a function of $m_K,\rho_n, \rho_p$ due to the kaon-nucleon interactions\footnote{In this section, we will denote baryonic number density by $\rho$ reserving $n$ for neutron.}:
 \be
 m_K^*= \omega(m_K,\rho_n, \rho_p, ...) \label{mko}
 \ee
 The density at which a neutron can decay into a proton and $K^-$ via the weak process, $n \rightarrow p + K^-$,
\be
\mu_n - \mu_p = m_K^*,
\ee
determines the condensed kaon amplitude $\rho_t$. Above the kaon condensation where $ m_K^*$ can be identified as the kaon chemical potential $\mu_K$, the chemical equilibrium is reached as
\be
\mu_n - \mu_p = \mu_e = \mu_{\mu} = \mu_K \equiv \mu.
\ee
where
\be
\mu_n - \mu_p&=& 4(1 - 2 \frac{\rho_p}{\rho})S(\rho) + \Theta(K)F(K,\mu) \label{betaeq0}
\ee
where $K$ stands for the kaon amplitude of kaon condensed state, i.e., $\la K\ra$,  and $F(K,\mu)$ is a nontrivial function that depends on the neutron-proton chemical potential difference which in turn depends on kaon-nucleon interactions. It is a highly model-dependent quantity but one can see how the kaon condensation threshold depends non-trivially on the nuclear symmetry energy, represented by the symmetry energy factor $S(\rho)$.

The Hamiltonian for the nucleons and negatively charged s-wave kaons\footnote{P-wave kaons couple to the nucleon to generate hyperons that we are ignoring here.} involved is taken in the simple form\cite{KLR}
\be
{\cal H} = {\cal H}_{KN}  + {\cal H}_{NN},
\ee
where
\be
{\cal H}_{KN} &=&  \partial_{0} K^- \partial^{0} K^+  + [m^2_K  -
\frac{n}{f^2} \Sigma_{KN}] K^+ K^-, \label{hkn}\\
{\cal H}_{NN} &=& \frac{3}{5}E_F^0 \left(\frac{\rho}{\rho_0}\right)^{2/3}
\rho + V(\rho) + \rho \left(1-2\frac{\rho_p}{\rho}\right)^2 S(\rho).\label{H}
\ee
Here $V(\rho)$ is the potential energy of nuclear matter as a function of density $\rho$ and $S(\rho)$ is the symmetry energy factor as a function of nuclear density. $\Sigma_{KN}$ is the KN sigma that represents the effect of the strange quark mass which is not zero (while the up and down quarks are taken as massless). It is defined as $\Sigma_{KN}\approx  \frac 12 (\bar{m}+m_s)\la N|\bar{u}s+\bar{s}s|N\ra$ and in medium, the nucleon matrix element of the bilinear quark fields will undergo medium modification. Therefore what it can be in (\ref{H}) is known neither theoretically nor experimentally. Even in matter-free space, the sigma term has evolved from $\sim 400$ MeV to $\sim 200$ MeV, the latest result coming from lattice QCD measurements. In \cite{KLR}, this uncertainty is taken into consideration.

The potential $V(\rho)$ is fit to nuclear matter density, so considered more or less known but beyond that density, it is highly model-dependent. What we will do is to pick one convenient parametrization called MDI (``momentum-dependent interaction")\footnote{e.g., the parametrization of Li {\it et al.}, Phys. Rept. {\bf 464}, 113 (2008).}. We shall see that while consistent within the error band with experimental constraints, they can give different results at higher density in the EoS when the strangeness and quark degrees of freedom are involved.

In terms of the amplitude of s-wave kaon condensation, $K$, and the kaon chemical potential, $\mu_K$, defined by the ansatz $K^{\pm} = K e^{\pm i \mu t}$, $F(K,\mu)$ in Eq.(\ref{betaeq0}) becomes
\be
F(K,\mu) = \frac{\mu}{2 f^2}K^2 +\cdots \label{FK}
\ee
Following the suggestion from the lattice QCD calculation mentioned above, ${\cal O} (K^{2n})$ terms for $n > 1$ figuring in the ellipsis are dropped. As density increases, the kaon chemical potential $\mu(=\mu_e)$ approaches 0 at the critical density.  Most of the electrons present there are converted to kaons which balance the charge neutrality of the system with the large proton fraction  $x=\frac{\rho_p}{\rho}=1/2$. Now if we consider this critical density as a phase boundary toward quark matter, then the chemical equilibrium (via confinement-deconfinement) reads
\be
\mu_n -\mu_p &=& \mu_d-\mu_u,\label{chemq} \\
\mu_{K^-} &=& \mu_s - \mu_u \label{chemk}
\ee
at the phase boundary. Note that the strange quark is required at the boundary, which implies that there should be a strange quark matter for $\rho > \rho_c$.
Since $\mu(=\mu_K) =0$ , we have from Eqs.~(\ref{chemq}) and (\ref{chemk})
\be
\mu_u ~ = ~ \mu_d \ \ {\rm and} \ \ \mu_s ~ = ~ \mu_u,
\ee
which gives
\be
\mu_u = \mu_d = \mu_s. \label{musqm}
\ee
This is the chemical potential relation for the SQM in the masselss limit.  In this simple picture, the kaon condensed nuclear matter  leads naturally to a strange quark matter.

\begin{figure}[t!]
\begin{center}
\includegraphics[width=8.6cm]{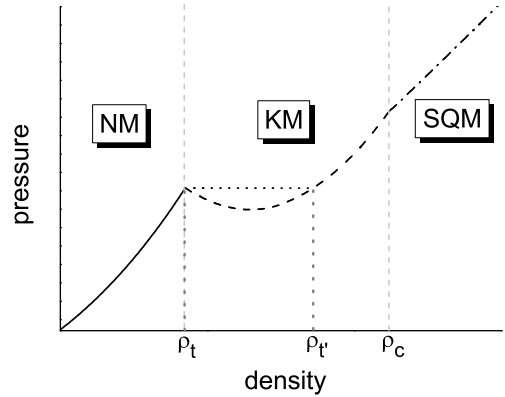}
\caption{The schematic phase diagram for NM-KNM-SQM.}
\label{nkqpre}
\end{center}
\end{figure}

For a choice of $\Sigma_{KN} \sim 260$ MeV which is compatible with current lattice results, we can have a triple-layered stellar structure consisting of NM, KNM and SQM from the  outer layer to the core part as shown schematically in Fig.~\ref{nkqpre}.
In this work,  we take a rather simple approach, namely, allow the discontinuity of density and chemical potential by assuming that NM with the density $\rho_t$  changes into KNM with  the density $\rho'_t$ at the phase boundary defined by
\be
P(\rho_t) = P(\rho'_{t}).
\ee
Assuming the strange quark matter to be in $SU(3)$ symmetric phase in the massless limit, we have the EoS of the form
\be
\epsilon_{SQM} &=& a_4\frac{9}{4\pi^2}\mu_q^4 + B ~~ = ~~ 4.83a_4\rho^{4/3} + B,\\
P_{SQM}        &=& a_4\frac{3}{4\pi^2}\mu_q^4 - B ~~ = ~~ 1.61a_4\rho^{4/3} - B,
\ee
where $B$ is the bag constant. Here $a_4$ denotes the perturbative  QCD correction\footnote{See A. Chodos et al., Phys. Rev. D {\bf 9}, 3471 (1974);
J.~M.~Lattimer and M.~Prakash, [arXiv:1012.3208 [astro-ph.SR]]; S.~Weissenborn, I.~Sagert, G.~Pagliara, M.~Hempel and J.~Schaffner-Bielich, [arXiv:1102.2869 [astro-ph.HE]].} which takes the value  $a_4 \leq 1$. The equality holds for SQM without QCD corrections.

\begin{figure}[t!]
\begin{center}
{\includegraphics[width=8.6cm]{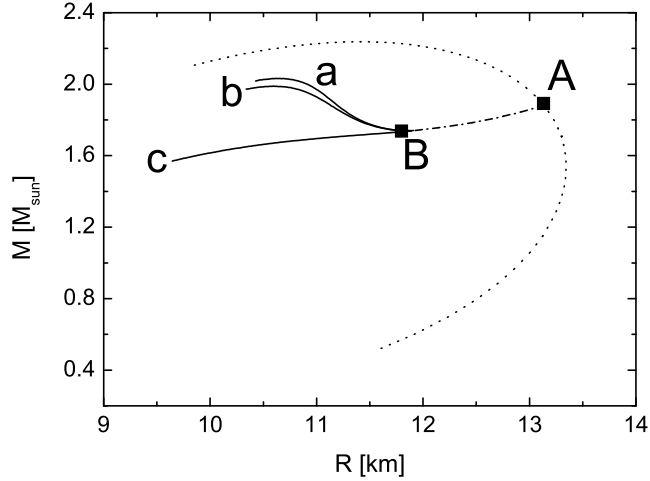}} \hfill
\caption{The M-R sequences for LCK with $\eta=-1$.  The dotted line denotes NM.  The dashed-dotted line between A and B denotes the double-layered (NM-KNM) system.  The solid lines, a, b and c,  denote the triple-layered (NM-KNM-SQM) system with the QCD corrections with $a_4 = 0.59, 0.62$ and $1$ respectively. }
\label{MR-}
\end{center}
\end{figure}

The resulting mass-radius relations that follow from the TOV equation are plotted in Fig.~\ref{MR-} using the parameter set, $\eta=-1$, $\Sigma_{KN}=259$ MeV and $B^{1/4}=97.5$MeV.   As expected, kaon condensation with higher $\Sigma_{KN}$'s would lead to smaller masses. However when the central density becomes higher, an SQM driven by kaons appears at the core. No sharp change due to the emergence of SQM is observed, which implies that the kaon driven SQM transition is a rather smooth transition in this scenario.  The maximum mass of $\sim 1.99 M\odot$ can be obtained by the gravitational instability condition  at the central density $\rho= 12.3\ \rho_0$. With $a_4=0.59$, a slightly larger maximum mass, $2.03 M\odot$, can be obtained, which is consistent with the recent observation.

Given the difficulty in calculating the symmetry energy for high density in
QCD-motivated approaches, it is tempting to try holographic techniques. Such a
trial has been made in \cite{sym-holography} which will be reviewed in Section 8.
In nuclear many-body approaches, short-range correlation functions tend to dampen the short-distance repulsion in nuclear forces, in particular in the tensor forces, thereby making the symmetry energy turn over and go down at $\rho\gsim 4\rho_0$. The suppression of the hard-core repulsion found in the dilaton limit discussed below corroborates this feature. We will show in Section \ref{hqcd} that holographic descriptions can make a simple and  unequivocal prediction that is different from what is given by nuclear many-body approaches. Whether or not this holographic QCD prediction is correct will perhaps be tested in future experiments.
\section{Dilaton Limit and the Fate of Repulsive Core}\label{dialton}
The ``soft" dilaton scalar figured in the mass scaling (\ref{BRS1}) in the HLS
nonlinear sigma model as discussed in
Section \ref{BR} can play an essential role in providing the attraction needed
in nuclear matter that saturates at the given density. The HLS
Lagrangian with baryons incorporated -- which is gauge equivalent to  baryonic
non-linear sigma model -- cannot however describe, in mean field, what happens
near the chiral phase transition. Near the transition point, the appropriate
Lagrangian is presumably more like the Gell-Mann-L\'evy (GML) linear sigma model. It is not known how density drives the dilatonic non-linear sigma model to the GML-type model. But this can be achieved {\em by fiat} by taking what is called ``dialton limit"\footnote{The dilaton limit used by us was discussed by S.R.  Beane and U. van Kolck in  Phys. Lett. {\bf B328}, 137 (1994).}.

In \cite{SLPR}, a parity-doublet hidden local symmetry Lagrangian with the soft dilaton suitably incorporated was constructed. The objective there was to
introduce the chiral-invariant fermion mass $m_0$ in the Lagrangian giving a
nucleon mass of the form $m_N=m_0 + C$ with $C$ that vanishes when
$\la\bar{q}q\ra\rightarrow 0$, so making the scaling (\ref{BRS2}) applicable. It has been shown that one can transform the dilaton baryonic HLS Lagrangian so constructed to the GML-type Lagrangian by taking the  dilaton limit
\be
g_{VN}\equiv g(1-g_V) &\rightarrow& 0,\label{gv}\\
g_A &\rightarrow & g_V,\label{ga}\\
m_N &\rightarrow&  m_0 \label{mass0}
\ee
where $g$ is the hidden gauge coupling, $g_{V,A}$ are the vector and axial-vector couplings in the Lagrangian and $g_{VN}$ is the effective vector-meson-nucleon coupling. The limit (\ref{mass0}) shows that at the dilaton-limit density $n_{\rm dilaton}$, the dropping of the nucleon mass will stop at $m_0$. That would account for the scaling (\ref{BRS2}).

\vskip 0.2cm
$\bullet$ {\bf Suppression of hard-core repulsion}
\vskip 0.2cm

The most dramatic prediction in the work of \cite{SLPR} is that as the density $n_{\rm dilaton}$ is approached, the vector coupling disappears. A potentially important consequence is that since the repulsive core -- in two-body as well as  multi-body forces -- is generated by $\omega$ exchanges in this model, the repulsion must disappear at high density as the vector mesons decouple. This would greatly modify the EoS of compact star matter, since the density in the interior of compacts stars can reach tens of nuclear matter density. What would be needed is to determine the density at which the suppression becomes effective and how it would influence the EoS.
\vskip 0.2cm
$\bullet$ {\bf Demise of mean-field theory at high density}
\vskip 0.2cm
In mean field theory, the symmetry energy goes as
\be
S\sim (g_{\rho N}^2/m_\rho^2) n.
\ee
Since in the dilaton limit, the $\rho NN$ coupling goes to zero faster than the $\rho$ mass subject to the vector manifestation (see (\ref{gv})), the symmetry energy should drop rapidly as density approaches the dilaton-limit density. This is at odds with the expression (\ref{tensorcon}) with the new BR. This implies that the mean-field approximation in HLS theory could go wrong at high density at least for the symmetry energy.
\vskip 0.2cm
$\bullet$ {\bf Dilaton limit is a fixed point}
\vskip 0.2cm

A natural question to raise is what is this ``dilaton limit"? Remarkably, we find by an RGE analysis~\cite{PLRS} that the dilaton limit exactly corresponds to the IR fixed point of HLS theory, which suggests that the system flows to that point as density is dialled. This fixed point is believed to sit slightly below the vector manifestation fixed point in hidden local symmetry theory at which the (hidden) gauge coupling itself goes to zero -- and the vector meson mass goes to zero.

\section{Symmetry Energy from hQCD}\label{hqcd}
Given that standard nuclear physics approaches give widely varying and uncontrolled results for the symmetry energy at high densities, it is highly appealing to resort to gravity/gauge duality which can in principle handle strongly-coupled phenomena like the EoS at high density. Here we give a brief summary of the effort made in that direction\cite{sym-holography}.

Based on the treatment of dense matter in confined phase  suggested in \cite{Seo:2008qc}, a simple model for nuclear matter to strange matter transition was proposed in \cite{KSS2010}, where two D6 branes for light and intermediate mass (strange) flavors were introduced.  The dense matter was introduced by uniformly distributed compact D4 branes with $N_c$ fundamental strings attached\footnote{Whether this uniform distribution is a realistic picture of dense matter is an open question and remains unsettled.}.
By considering energy minimization, transition from nuclear to strange matter could be studied. To calculate the symmetry energy in nuclear matter, we consider the case where the two flavors have the same quark masses, $m_1=m_2$.  We find that the symmetry energy increases monotonically with the total charge $Q$ which is roughly proportional to the square-root of the density.

The model we use to study the symmetry energy is the D4/D6/D6 model~\cite{KSS2010} with  baryon vertices consisting of D4 branes and fundamental strings. In our approach, gluon dynamics is replaced by the gravity sourced by the $N_c$-colored D4 branes, and two probe D6 branes are used to describe the up and down quarks. The bare quark masses are the distances between the D4 and the two D6's in the absence of string coupling.
We wrap the D4 brane on $S^4$ which is transverse to the original D4 brane.
Due to the Chern-Simons interaction with RR-field, a $U(1)$ gauge field is induced on the D4 brane world volume.

The source of the gauge field is interpreted as the end point of fundamental strings. Substituting the equation of motion for gauge field into the
Dirac-Born-Infeld action of D4 brane with the Chern-Simons term, we get `Hamiltonian' for the D4 brane. The other endpoints of fundamental strings  are attached to two D6 branes with a given number ratio. The endpoints of fundamental strings provide the source of the $U(1)$ gauge field on the D6 brane. The two D6 branes are connected to a D4 brane by fundamental strings.
Therefore, the D6 branes are pulled down and the compact D4 brane is pulled up.

\begin{figure}[!ht]
\begin{center}
\includegraphics[angle=0, width=0.5\textwidth]{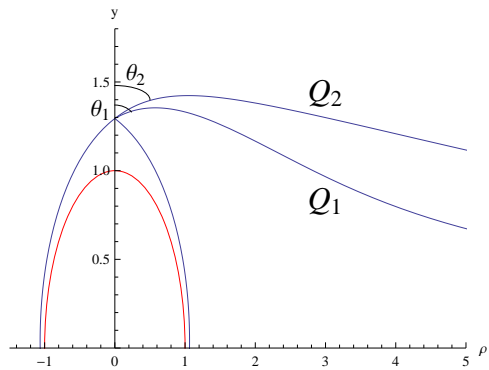}
\caption{(Color online) Embedding of D-branes with $\alpha \ne 0.5$. The asymptotic heights of two branes are the same ($m_1=m_2 =0.1$). The red curve denotes the position of $U_{KK}$. } \label{config}
\end{center}
\end{figure}

As discussed in \cite{Seo:2008qc}, the length of the fundamental strings becomes zero since the tension of the fundamental strings is always larger than that of D-branes. Finally, the position of the cusp of D6 branes should be located at the same position of the cusp of the D4 brane, $\xi_c$. We consider $Q_1$ fundamental strings attached to one D6 brane and $Q_2$ strings attached to another D6 brane. The force at the cusp of the D6 branes must be balanced by the pulling of the baryon vertex D4. See the figure \ref{config} for the final configuration of the baryon vertex and the probe flavor D6 brane.

\begin{figure}[!ht]
\begin{center}
\includegraphics[angle=0, width=0.6\textwidth]{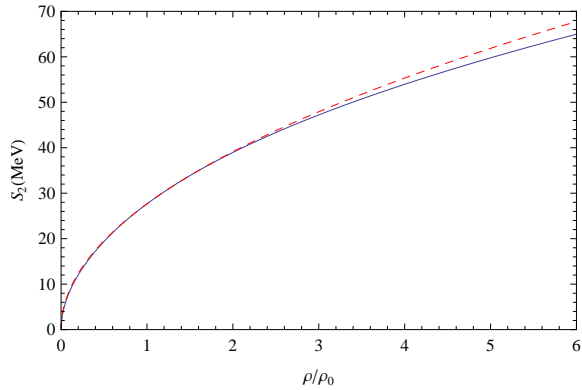}
\caption{(Color online) Solid line is the   symmetry energy as a function of density.
The  dotted line is the best fit of
  $S_2$ with $ \rho^{1/2}$.  }\label{symmetryE1}
\end{center}
\end{figure}

For the nuclear symmetry energy, we consider only the  $m_2/m_1=1$ case.
The explicit form of the symmetry energy per nucleon can be written as
\be\label{Es}
S_2 =\frac{2\tau_6}{N_B} \int d\rho \frac{\sqrt{1+\dot{y}^2}\tilde{Q}^2
\omega_+^{10/3} \rho^4}{(\tilde{Q}^2 +4 \omega_+^{8/3} \rho^4)^{3/2}},
\ee
where $y$ is the embedding solution of D6 brane with $\tilde{\alpha}=0$.
Since $N_B=Q/N_c$, the symmetry energy (\ref{Es}) contains an $N_c$ factor.
We need to factor this $N_c$ out for the reason  we discuss later.
Our results are given in Fig. \ref{symmetryE1}. Note that so far we have used $\rho$ for both the coordinate and the density. Hereafter $\rho$ will denote  only the density.\footnote{We continue to denote density by $\rho$ as in Section \ref{star}.} To fix the energy scale, we used the value of the 't Hooft coupling $\lambda$ and the compactification scale $M_{KK}$ determined in \cite{JKS}.
We stress that there are two notable aspects in our results that are rather insensitive to the choice of $\lambda$ and $M_{KK}$. One is the stiffness of the symmetry energy $S_2$ in the high density regime, and the other is its low density power-law  behavior $S_2\sim \rho^{1/2}$.

The power-law behavior of $S_2$ in the low density regime can be understood
by calculating  analytically  in a special limit, $m_q \rightarrow \infty$ and $\rho \rightarrow 0$. In this case, the solution of the D6 brane embedding becomes trivial, $\dot{y}=0$, and we can integrate (\ref{Es}) analytically to have
\be
S_2 =\left(\Gamma(\frac{5}{4})\right)^2\sqrt{\frac{ \lambda \rho_0}{2 M_{KK}}} \sqrt{\frac{\rho}{\rho_0}}.
\ee
The current experimental result  of the symmetry energy
can be summarized by a fitting formula
\bea
S_2 (\rho)=c(\rho/\rho_0)^\gamma
\eea
 with $c\simeq31.6$ MeV and $\gamma=0.5-0.7$
in the low density regime, $0.3\rho_0\le \rho\le \rho_0$.\footnote{See, e.g.,   D.V. Shetty and S.J. Yennello, Pramana {\bf 75} 259 (2010).}
With  our choice of $\lambda, M_{KK}$, we obtain $\gamma\simeq 0.5$ and $c\simeq 27.7$ MeV. We note that the value of $\gamma$  in our results is rather insensitive to the value of $\lambda$ and $M_{KK}$, while
the value of $c$ depends on them.

In standard nuclear physics approaches, at very low densities $\rho\ll \rho_0$, the dominant contribution to the symmetry energy comes from the kinetic energy term which encodes the  Pauli principle. This is because the kinetic contribution to the symmetry energy is $\sim\rho^{2/3}$, while the one from interactions starts from $\sim \rho^1$ due to the linear density approximation which works well at very low density. The origin of the exponent  $\gamma=2/3$ is the dispersion relation $E\sim p^2$ together with the sharp Fermi surface.
In our case, the fact that $\gamma=1/2$  suggests that either the dispersion relation is anomalous like $E\sim p^{3/2}$ or
Fermi surface is fuzzy.\footnote{As in condensed matter discussed by, e.g., S.~S.~Lee,   Phys.\ Rev.\  D {\bf 79}, 086006 (2009) and
  T.~Faulkner {\it et al.},  arXiv:0907.2694 [hep-th].}
This is a striking novel prediction of hQCD, not anticipated in nuclear field theory or many-body approaches and poses an interesting future study.

\section{Role of the Infinite Tower of Vector Mesons in Nucleon Structure}  Although in all of the phenomena discussed above, hidden local symmetry is seen  to play a significant or even crucial role, it is {\em not} a fundamental symmetry of QCD. It is in some sense an emergent (``hidden") symmetry arising from collective excitations of the  relevant degrees of freedom,  i.e., pions, quasiquarks or nucleons etc. Up to date, only  the lowest-lying vector mesons $\rho$, $\omega$ have been considered in the gauge symmetric framework. This is because the scale considered is the chiral  scale $2\pi f_\pi \sim 1$ GeV.

Considered as an emergent symmetry, it is natural to expect that as the energy scale is increased, more hidden gauge particles could appear. In fact, the recent development of holographic QCD models indicates that an infinite tower of vector mesons can provide a more powerful framework than just limiting to the lowest excitations. An interesting question that is raised in this program is how the infinite tower of vector mesons affect the structure of dense matter.

As the first step toward this goal, we consider the nucleon electromagnetic
(EM) form factors. It is well-known that the famous Sakurai vector dominance
picture of 1960's works quantitatively well for the pion form factor but fails
completely for the nucleon form factors. It turns out that this defect for the
latter can be remedied in holographic QCD in terms of a 5D Yang-Mills action in the bulk. The vector dominance is found to work equally well for both the pion
and the nucleon when the infinite tower, natural in holographic QCD, is included.
 \begin{figure}[hbtp]
 \begin{center}
 \includegraphics[width=9cm]{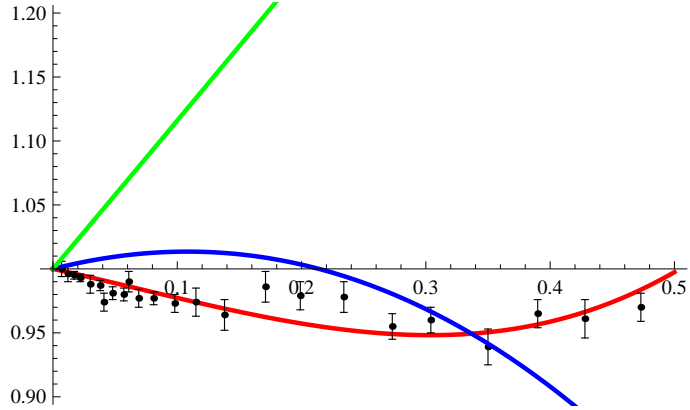}
 \end{center}
 \caption[]{(Color online) $G_E^p/G_D$ vs. $Q^2$.  Here $G_D$ is the well-known dipole form  factor. Green: Sakurai VD, Blue: hQCD prediction, Red: best fit}\label{GE}
 \end{figure}

 A simple but highly quantitative way to see this at low-momentum transfers is
to integrate out all higher members of the tower leaving only the lowest vector mesons $V= (\rho,\omega)$ active and write the resulting form factors in terms of $V$ in a hidden local symmetric form. The physics of the high tower will then be captured in the parameters that figure in the resulting HLS Lagrangian. Based on the consideration of chiral perturbation theory with hidden local fields, one can write, say, for the proton electric Sachs form factor for low momentum transfers $Q^2 < 1$ GeV$^2$ as~\cite{HR-integrate}
 \be
 G_{E}^p (Q^2)
 = \left( 1 - \frac{a_{E}}{2} \right) + z_{E}  \frac{Q^2}{m_\rho^2} +
 \frac{a_{E}}{2} \, \frac{m_\rho^2}{m_\rho^2 + Q^2 }+\cdots
 \ ,
 \label{ge}
 \ee
 where $a_E$ and $z_E$ are parameters given by the theory. The expression (\ref{ge}) can be understood as the chiral perturbation expression calculated to ${\cal O}(p^4)$ with the vector meson treated on the same footing as the pion. The third term with the vector-meson propagator is a specific feature characterizing the fact that the $\rho$ meson and the pion are treated on the same footing, that is, they are of the same scale. Thus perturbative unitarity is applied both to the pion and the $\rho$, given in terms of the propagator which corresponds to the infinite sum of the chiral expansion. This is the unique feature of hidden local symmetry theory not shared by other treatments where the $\rho$ mass is considered as ``heavy" compared with the pion mass.

The second term is an ${\cal O}(p^4)$ contribution which is leading order in $N_c$ counting. The pion loop terms contribute to the same chiral order but they are down by $1/N_c$, so can be ignored in the large $N_c$ counting.

In the form given, the Sakurai VD corresponds to setting
$a_E=2$ and $z_E=0$. It is given by the green curve in
Fig.~\ref{GE}. It shows in a spectacular way what we have known since a long time, that is, it fails very badly. Holographic QCD models with infinite tower would predict what's given in blue which is not so bad for the electric form factor. A best fit, which is what would be expected for the holographic QCD models if $1/N_c$ corrections were suitably taken into account, is given in red. It agrees well with Nature, with $\chi^2/{\rm dof}\sim 3/2$. It shows the possibly important role of the infinite tower in the nucleon structure and more significantly, in dense baryonic matter.

It turns out~\cite{HR-integrate} that the Sakai-Sugimoto holographic model which is known to be the only holographic model with the chiral symmetry of QCD in the chiral limit does not fare well with the proton magnetic moment. This is not difficult to understand. In QCD, $1/N_c$ corrections are expected to be relatively insignificant in the electric form factor but cannot be ignored in the magnetic form factor. The SS model lacks such $1/N_c$ corrections.
\section{Comments and Projects}
We have formulated a rather unconventional approach to dense compact-star
matter starting with hidden local symmetric Lagrangian in which the dilaton
degree of freedom associated with the trace anomaly of QCD and the topological
solitons as baryonic degrees of freedom are incorporated.  Some novel
phenomena, hitherto neither observed experimentally nor predicted
theoretically, have been discovered, but their validity remains up to date mostly untested. In anticipation of the forthcoming accelerators such as RIB machines (e.g., KoRIA), FAIR/GSI, J-PARC etc. and to  confront the data that will come from them, we need to sharpen the arguments that so far have been somewhat short in rigor and make the calculations more quantitative and precise. In doing so, we would like to address the following questions:
\begin{itemize}
\item How to systematically and accurately formulate nuclear matter,
    strange matter and quark matter so as to map out the phase diagram of
    dense baryonic matter in the strategy formulated in the program as
    described above?  This aims to solidify the prediction -- and fit to
    the data on the 1.97 solar mass star -- made in
    \cite{KLR}.

\item How does the new BR scaling affect nuclear structure, nuclear matter, neutron-rich nuclei and the EoS of neutron stars?\cite{Dongetal}  This issue is closely related to what will be measured by RIB machines (e.g., KoRIA) and FAIR/GSI.

\item What is the role of the infinite tower in the short-distance properties of nuclear interactions along the line that was discussed in \cite{HR-integrate}, e.g., the nucleon core size etc., and in dense baryonic matter?~\cite{haradaetal}

\item How does the infinite tower structure of skyrmions in holographic QCD affect the deeply bound kaonic nuclei and kaon condensation in dense
    baryonic matter?~\cite{haradaetal}
\item How to exploit the power of gravity dual theory in the highly
    nonperturbative density regime relevant to compact stars that cannot be accessed by QCD?
\item What are the specific predictions made in our approach in the density regime that can be accessed by the forthcoming RIB machines?
\end{itemize}

Closely tied to the main theme underlying the above issues, there is
the possibility to probe dense matter EoS via gravity wave. This matter will be explored in the years to come~\cite{KL-gravity}.

\subsection*{Acknowledgments}

The work reported here was partially supported by the WCU project of Korean
Ministry of Education, Science and Technology (R33-2008-000-10087-0).  We are
grateful for discussions and collaborations with Masa Harada, Kwanghyun Jo,
Youngman Kim, Kyungmin Kim, Tom Kuo, Yongseok Oh, Won-Gi Paeng, Byung-Yoon Park, Chihiro Sasaki and Ismail Zahed.


\end{document}